\documentclass[english]{article}
\usepackage[T1]{fontenc}
\usepackage[latin1]{inputenc}
\usepackage{setspace}
\usepackage{amssymb}

\makeatletter

\newcommand{\lyxaddress}[1]{
\par {\raggedright #1
\vspace{1.4em}
\noindent\par}
}

\usepackage{babel}
\makeatother
\begin{document}

\title{Generalized Electromagnetic fields of Dyons in Inhomogenous Media}

\author{P. S. Bisht, O. P. S. Negi and Jivan Singh }

\maketitle
\begin{singlespace}

\lyxaddress{\begin{center}Department of Physics\\
 Kumaun University\\
 Soban Singh Jeena Campus\\
 Almora-263601 (Uttarakhand) \\
INDIA\par\end{center}}

\lyxaddress{\begin{center}Email:- ps\_bisht123@rediffmail.com\\
 ops\_negi@yahoo.co.in\\
jivan@indiatimes.com\par\end{center}}
\end{singlespace}

\begin{abstract}
Reformulation of the generalized electromagnetic fields of dyons has
been dicussed in inhomogenous media and corresponding quaternionic
equations are derived in compact, simple and unique manner. We have
also discussed the monochromatic fields of generalized electromagnetic
fields of dyons in slowly changing media in a consistent manner.
\end{abstract}
In the series of papers we have discussed the Maxwell's-Dirac equation
in the homogenous (isotropic) medium and its solutions \cite{key-1},
quaternionic form of the generlaized fields of dyons in homogenous
(isotropic) medium and other quantum equations \cite{key-2}, time-harmonic
form of generalized fields of dyons in homogenous (isotropic) medium
\cite{key-3} and in the presence of source the time-dependent Maxwell's
-Dirac equation have been developed in chiral media and the solution
for the problem are obtained in unique and consistent manner \cite{key-4}.
In this paper we have reformulated the generalized electromagnetic
fields of dyons in inhomogenous media and corresponding quaternionic
equations are derived in compact, simple and unique manner. Finally
we have also discussed the monochromatic fields of generalized electromagnetic
fields of dyons in slowly changing media in a consistent manner.

Postulating the existence of magnetic monopoles\cite{key-5}, let
us write the Generalized Dirac Maxwell (GDM) field equations in isotropic
(homogenous) medium in the following manner \cite{key-1} in terms
of natural units $c=\hbar=1$;

\begin{eqnarray*}
\overrightarrow{\nabla}.\overrightarrow{D} & = & \rho_{e}\end{eqnarray*}
\begin{eqnarray*}
\overrightarrow{\nabla}.\overrightarrow{B} & = & \mu\rho_{m}\end{eqnarray*}

\begin{eqnarray*}
\overrightarrow{\nabla}\times\overrightarrow{E} & = & -\frac{\partial\overrightarrow{B}}{\partial t}-\frac{\overrightarrow{j_{m}}}{\epsilon}\end{eqnarray*}

\begin{eqnarray}
\overrightarrow{\nabla}\times\overrightarrow{H} & = & \frac{\partial\overrightarrow{D}}{\partial t}+\overrightarrow{j_{e}}\label{eq:1}\end{eqnarray}
Where $\rho_{e}$ and $\rho_{m}$ are respectively the electric and
magnetic charge densities, $\overrightarrow{j_{e}}$ and $\overrightarrow{j_{m}}$
are the electric and magnetic current densities, $\overrightarrow{D}$
is electric displacement vector, $\overrightarrow{E}$ is the electric
field, $\overrightarrow{B}$ is magnetic induction vector and $\overrightarrow{H}$
is magnetic field. $\varepsilon$ and $\mu$ are respectively the
permiability and permitivity of the media.

Assuming that $\epsilon$ and $\mu$ are the function of coordinates
\cite{key-6} such as

\begin{eqnarray*}
\epsilon & = & \epsilon(x)\end{eqnarray*}

and

\begin{eqnarray}
\mu & = & \mu(x).\label{eq:2}\end{eqnarray}
The Maxwell's equations are thus considered together with the relation
given by (\ref{eq:2}) may then be written accordingly in inhomogeneous
media describing relations among the induction and the field vectors.
Let us now define $\overrightarrow{D}$ and $\overrightarrow{B}$
as,

\begin{eqnarray*}
\overrightarrow{D} & = & \overrightarrow{D}\,(\overrightarrow{E},\overrightarrow{H})\end{eqnarray*}

\begin{eqnarray}
\overrightarrow{B} & = & \overrightarrow{B}\,(\overrightarrow{E},\overrightarrow{H})\label{eq:3}\end{eqnarray}
and

\begin{eqnarray*}
\overrightarrow{D} & = & \epsilon_{0}\epsilon_{r}\overrightarrow{E}\end{eqnarray*}

\begin{eqnarray}
\overrightarrow{B} & = & \mu_{0}\mu_{r}\overrightarrow{H}.\label{eq:4}\end{eqnarray}
Using the relation (\ref{eq:3}), we may write the generalized Dirac-Maxwell
(GDM) equations (\ref{eq:1}) for the isotropic inhomogenous medium
for dyons as

\begin{eqnarray*}
\overrightarrow{\nabla}.(\epsilon\overrightarrow{E}) & = & \rho_{e}\end{eqnarray*}

\begin{eqnarray*}
\overrightarrow{\nabla}.\overrightarrow{(B} & )= & \mu\rho_{m}\end{eqnarray*}

\begin{eqnarray*}
\overrightarrow{\nabla}\times\overrightarrow{E} & = & -\frac{\partial\overrightarrow{B}}{\partial t}-\frac{\overrightarrow{j_{m}}}{\epsilon}\end{eqnarray*}

\begin{eqnarray}
\overrightarrow{\nabla}\times\overrightarrow{B} & = & \frac{1}{v^{2}}\frac{\partial\overrightarrow{E}}{\partial t}+\mu\overrightarrow{j_{e}}.\label{eq:5}\end{eqnarray}
The first and second differential equations of GDM equations (\ref{eq:5})
can be written as follows \cite{key-7} ,

\begin{eqnarray}
div\overrightarrow{E} & +<\frac{grad\epsilon}{\epsilon},\,\,\, & \overrightarrow{E}>=\frac{\rho_{e}}{\epsilon}\label{eq:6}\end{eqnarray}

and

\begin{eqnarray}
div\overrightarrow{B} & +<\frac{grad\mu}{\mu},\,\,\, & \overrightarrow{B}>=\mu\rho_{m}.\label{eq:7}\end{eqnarray}
Combining equations (\ref{eq:6}) and (\ref{eq:7}) with the third
and fourth equation of (\ref{eq:5}), we obtain the following form
of generalized Dirac-Maxwell (GDM) equations i.e.

\begin{eqnarray}
div\overrightarrow{E} & =<\frac{grad\epsilon}{\epsilon},\,\,\, & \overrightarrow{E}>-\frac{\partial\overrightarrow{B}}{\partial t}-\frac{\overrightarrow{j_{m}}}{\epsilon}-\frac{\rho_{e}}{\epsilon}\label{eq:8}\end{eqnarray}
and

\begin{eqnarray}
div\overrightarrow{B} & =<\frac{grad\mu}{\mu},\,\,\, & \overrightarrow{B}>\frac{1}{v^{2}}\frac{\partial\overrightarrow{E}}{\partial t}-\mu\overrightarrow{j_{e}}-\mu\rho_{m}.\label{eq:9}\end{eqnarray}
Let us take the scalar products of two vectors $\overrightarrow{p}$
and $\overrightarrow{q}$ as,

\begin{eqnarray}
<\overrightarrow{p},\overrightarrow{q}> & = & -\frac{1}{2}(^{\overrightarrow{p}}M+M^{\overrightarrow{p}})\overrightarrow{q}.\label{eq:10}\end{eqnarray}
Using this equation (\ref{eq:10}), we get the following pair of equations
from equations (\ref{eq:8}) and (\ref{eq:9}) as \begin{eqnarray}
(D+\frac{1}{2}\,\frac{grad\epsilon}{\epsilon})\,\overrightarrow{E} & =-\frac{1}{2}M^{\frac{grad\epsilon}{\epsilon}}\overrightarrow{E}\,-\frac{\partial\overrightarrow{B}}{\partial t}-\frac{\overrightarrow{j_{m}}}{\epsilon}- & \frac{\rho_{e}}{\epsilon}\label{eq:11}\end{eqnarray}
and

\begin{eqnarray}
(D+\frac{1}{2}\,\frac{grad\mu}{\mu})\,\overrightarrow{B} & =-\frac{1}{2}M^{\frac{grad\mu}{\mu}}\overrightarrow{B}\,- & \frac{1}{v^{2}}\frac{\partial\overrightarrow{E}}{\partial t}-\mu\overrightarrow{j_{e}}-\mu\rho_{m}.\label{eq:12}\end{eqnarray}
where we have used the following subsidiary condition

\begin{eqnarray}
\frac{1}{2}\,\frac{grad\epsilon}{\epsilon} & = & \frac{grad\sqrt{\epsilon}}{\sqrt{\epsilon}}.\label{eq:13}\end{eqnarray}
Then using the condition given by Kravehenko \cite{key-7}, we may
write equations (\ref{eq:11}) and (\ref{eq:12}) as follows,

\begin{eqnarray}
\frac{1}{\sqrt{\epsilon}}D(\sqrt{\epsilon}.\overrightarrow{E})+\overrightarrow{E}.\overrightarrow{\epsilon} & = & =-\frac{\partial\overrightarrow{B}}{\partial t}-\frac{\overrightarrow{j_{m}}}{\epsilon}-\frac{\rho_{e}}{\epsilon}\label{eq:14}\end{eqnarray}

\begin{eqnarray}
\frac{1}{\sqrt{\mu}}D(\sqrt{\mu}.\overrightarrow{B})+\overrightarrow{B}.\overrightarrow{\mu} & = & \frac{1}{v^{2}}\frac{\partial\overrightarrow{E}}{\partial t}-\mu\overrightarrow{j_{e}}-\mu\rho_{m}\label{eq:15}\end{eqnarray}
where

\begin{eqnarray*}
\overrightarrow{\epsilon} & = & \frac{grad\sqrt{\epsilon}}{\sqrt{\epsilon}};\end{eqnarray*}
and\begin{eqnarray*}
\overrightarrow{\mu} & = & \frac{grad\sqrt{\mu}}{\sqrt{\mu}}.\end{eqnarray*}
Rearranging the equations (\ref{eq:14}) and (\ref{eq:15}), we get
the following new set of equations i.e.

\begin{eqnarray}
(D+M^{\overrightarrow{\epsilon}})\overrightarrow{E} & = & -\frac{\partial\overrightarrow{B}}{\partial t}-\frac{\overrightarrow{j_{m}}}{\epsilon}-\frac{\rho_{e}}{\epsilon}\label{eq:16}\end{eqnarray}

\begin{eqnarray}
(D+M^{\overrightarrow{\mu}})\overrightarrow{B} & = & \frac{1}{v^{2}}\frac{\partial\overrightarrow{E}}{\partial t}-\mu\overrightarrow{j_{e}}-\mu\rho_{m}.\label{eq:17}\end{eqnarray}
In order to write the quaternionic equation for electromagnetic fields
in inhomogeneous medium, let us start with the following representations
for the quaternionic differential operator \cite{key-2},

\begin{eqnarray}
\boxdot & = & (\frac{i}{v}\partial_{t}+D)\label{eq:18}\end{eqnarray}
and its conjugate as

\begin{eqnarray}
\overline{\boxdot} & = & (-\frac{i}{v}\partial_{t}-D)\label{eq:19}\end{eqnarray}
where $v=\frac{1}{\sqrt{\epsilon\mu}}$ is the speed of the electromagnetic
wave in the medium. We define the complex vector field $\overrightarrow{\psi}$
associated with the generalized electromagnetic fields of dyons as
\cite{key-2},

\begin{eqnarray}
\overrightarrow{\psi} & = & \overrightarrow{E}-i\, v\,\overrightarrow{B}.\label{eq:20}\end{eqnarray}
Accordingly,  we can express the quantum equations associated with
the generalized four-current, electric and magnetic fields of dyons
in quaternionic form as,

\begin{eqnarray}
j & = & -i\rho v+j_{1}e_{1}+j_{2}e_{2}+j_{3}e_{3}\label{eq:21}\end{eqnarray}

\begin{eqnarray}
E & = & E_{1}e_{1}+E_{2}e_{2}+E_{3}e_{3}\label{eq:22}\end{eqnarray}

\begin{eqnarray}
B & = & B_{1}e_{1}+B_{2}e_{2}+B_{3}e_{3}\label{eq:23}\end{eqnarray}

\begin{eqnarray}
D & = & D_{1}e_{1}+D_{2}e_{2}+D_{3}e_{3}\label{eq:24}\end{eqnarray}
where $e_{0},\, e_{1},\, e_{2},\, e_{3}$ are the quaternion units
and satisfy the following commutation rules,

\begin{eqnarray}
e_{i}e_{j} & = & -\delta_{ij}+\varepsilon_{ijk}e_{k}\label{eq:25}\end{eqnarray}
where $\delta_{ij}$ is Kronecker delta and $\varepsilon_{ijk}$ is
the Levi-Civita three index symbol.

Operating equation (\ref{eq:19}) to equation (\ref{eq:20}) and using
equations (\ref{eq:16}) and (\ref{eq:25}), we get

\begin{eqnarray}
\overline{\boxdot}\psi & = & (M^{\overrightarrow{\epsilon}}\overrightarrow{E}+ivM^{\overrightarrow{\mu}}\overrightarrow{B})+iv\mu(-iv\overline{\rho}+j)\label{eq:26}\end{eqnarray}
where $j$ is the generalized current density of dyon and given by,

\begin{eqnarray}
j & = & j_{e}-i\, v\, j_{m}\label{eq:27}\end{eqnarray}
and $\overline{\rho}$ is the conjugate of the generalized charge
density of dyon and defined as,

\begin{eqnarray}
\rho & = & \rho_{e}-i\frac{\rho_{m}}{v}.\label{eq:28}\end{eqnarray}
From equation (\ref{eq:20}), we get the following relations,

\begin{eqnarray}
\overrightarrow{E} & = & \frac{1}{2}(\psi+\psi*)\label{eq:29}\end{eqnarray}
and

\begin{eqnarray}
\overrightarrow{B} & = & \frac{1}{2iv}(\psi*-\psi)\label{eq:30}\end{eqnarray}
where $\psi*$ is the complex conjugate of $\psi$ .Taking the first
part of the right hand side of equation (\ref{eq:26}), we get

\begin{eqnarray}
M^{\overrightarrow{\epsilon}}\overrightarrow{E}+ivM^{\overrightarrow{\mu}}\overrightarrow{B} & = & \frac{1}{2}(M^{(\overrightarrow{\epsilon}-\overrightarrow{\mu})}\psi+M^{(\overrightarrow{\epsilon}+\overrightarrow{\mu})}\psi*).\label{eq:31}\end{eqnarray}
We may now introduce the notation

\begin{eqnarray}
\overrightarrow{\epsilon}+\overrightarrow{\mu} & = & -\frac{gradv}{v}\label{eq:32}\end{eqnarray}
and

\begin{eqnarray}
\overrightarrow{\epsilon}-\overrightarrow{\mu} & = & -\frac{gradW}{W}\label{eq:33}\end{eqnarray}
where $W$ is the intrinsic wave impedance of the medium. The $\overrightarrow{v}$
and $\overrightarrow{W}$ are also expressed as

\begin{eqnarray}
\overrightarrow{v}: & = & \frac{grad\sqrt{v}}{\sqrt{v}}\label{eq:34}\end{eqnarray}
and

\begin{eqnarray}
\overrightarrow{W}: & = & \frac{grad\sqrt{W}}{\sqrt{W}}.\label{eq:35}\end{eqnarray}
Thus equation (\ref{eq:31}) reduces to

\begin{eqnarray}
M^{\overrightarrow{\epsilon}}\overrightarrow{E}+i\, v\, M^{\overrightarrow{\mu}}\,\overrightarrow{B} & = & -(M^{\overrightarrow{v}}\overrightarrow{\psi}*+M^{\overrightarrow{W}}\overrightarrow{\psi})\label{eq:36}\end{eqnarray}
As such, from equation (\ref{eq:26}), we obtain the generalized Dirac
Maxwell (GDM) equation of dyons for an inhomogenous medium as under,

\begin{eqnarray}
\overline{\boxdot}\psi+ & (M^{\overrightarrow{v}}\overrightarrow{\psi}*+M^{\overrightarrow{W}}\overrightarrow{\psi} & )=iv\mu(-iv\overline{\rho}+j)=i\, v\,\mu\,\overline{j}.\label{eq:37}\end{eqnarray}
So, the equation (\ref{eq:37}) is completely analogous to the generalized
Dirac Maxwell (GDM) equation of dyons given by equation (\ref{eq:5})
and represents the quaternionoc form of Maxwell's equation for dyons
in inhomogenous medium.

Let us define a time harmonic electromagnetic field as \cite{key-3},

\begin{eqnarray}
\overrightarrow{E}(x,t) & = & Re(\overrightarrow{E}(x)e^{-i\omega t})\label{eq:38}\end{eqnarray}
and

\begin{eqnarray}
\overrightarrow{B}(x,t) & = & Re(\overrightarrow{B}(x)e^{-i\omega t}).\label{eq:39}\end{eqnarray}
Here $\omega$ is the frequency of oscillation. We may now assume
that the sources are also considered as time-harmonic and accordingly
we may write the charge and current source densities of dyons as

\begin{eqnarray}
\rho(x,t) & = & Re(\rho(x)e^{-i\omega t})\label{eq:40}\end{eqnarray}
and

\begin{eqnarray}
\overrightarrow{j}(x,t) & = & Re(\overrightarrow{j}(x)e^{-i\omega t}).\label{eq:41}\end{eqnarray}
Hence for the case when electromagnetic field is considered to be
monochromatic, we may obtain the following sets of quaternionic equations
for generalized fields of dyons on substituting equations (\ref{eq:38},\ref{eq:39},\ref{eq:40},\ref{eq:41})
into the equations (\ref{eq:16}) and (\ref{eq:17}) i.e.

\begin{eqnarray}
D_{\overrightarrow{\epsilon}}\overrightarrow{E} & = & i\omega\overrightarrow{B}-\frac{\rho_{e}}{\epsilon}-\frac{\overrightarrow{j_{m}}}{\epsilon}=i\alpha v\overrightarrow{B}-\frac{\rho_{e}}{\epsilon}-\frac{\overrightarrow{j_{m}}}{\epsilon}\label{eq:42}\end{eqnarray}
and

\begin{eqnarray}
D_{\overrightarrow{\mu}}\overrightarrow{B} & = & -\frac{i\omega}{v^{2}}\overrightarrow{E}-\mu\rho_{m}+\mu\overrightarrow{j_{e}}=-\frac{i\alpha}{v}\overrightarrow{E}-\mu\rho_{m}+\mu\overrightarrow{j_{e}}.\label{eq:43}\end{eqnarray}
In equation (\ref{eq:42}) and (\ref{eq:43}) the quantity $\alpha=\frac{\omega}{v}$
is described as the wave number. Hence for sourceless condition, the
equation (\ref{eq:42}) and (\ref{eq:43}) are reduced to

\begin{eqnarray}
D_{\overrightarrow{\epsilon}}\overrightarrow{E} & = & i\alpha v\overrightarrow{B}\label{eq:44}\end{eqnarray}
and

\begin{eqnarray}
D_{\overrightarrow{\mu}}\overrightarrow{B} & = & -\frac{i\alpha}{v}\overrightarrow{E}.\label{eq:45}\end{eqnarray}
The medium is said to be slowly changing when its properties change
appreciably over distances much greater than the wavelength \cite{key-8}.
Generally that is associated with the possibilty of reducing the generalized
Dirac-Maxwell (GDM) equations of dyons (\ref{eq:44}) and (\ref{eq:45})
to the following forms of Helmholtz equations,

\begin{eqnarray}
(\bigtriangleup+\alpha^{2})\overrightarrow{E} & = & 0\label{eq:46}\end{eqnarray}
and

\begin{eqnarray}
(\bigtriangleup+\alpha^{2})\overrightarrow{B} & = & 0\label{eq:47}\end{eqnarray}
where

\begin{eqnarray}
\bigtriangleup+\alpha^{2} & =-(D+\alpha)(D-\alpha)= & -D_{\alpha}D_{-\alpha}.\label{eq:48}\end{eqnarray}
For checking the reduction , we consider that $|\overrightarrow{\epsilon}|$
and $|\overrightarrow{\mu}|$ are very small and the terms containing
the vectors $\overrightarrow{\epsilon}$ and $\overrightarrow{\mu}$
are supposed to be negligible. Then equations (\ref{eq:44}) and (\ref{eq:45})
are reduced to the following simple and compact quaternionic forms
i.e.

\begin{eqnarray}
D\overrightarrow{E} & = & i\alpha v\overrightarrow{B}\label{eq:49}\end{eqnarray}
and

\begin{eqnarray}
D\overrightarrow{B} & = & -\frac{i\alpha}{v}\overrightarrow{E}\label{eq:50}\end{eqnarray}
These equations can be diagonalized. Hence we may obtain the following
compact quaternionic equations for the functions $\psi$ and $\psi*$
of dyons i.e.

\begin{eqnarray}
D_{-\alpha}\overrightarrow{\psi} & = & 0\label{eq:51}\end{eqnarray}

and

\begin{eqnarray}
D_{\alpha}\overrightarrow{\psi}* & = & 0.\label{eq:52}\end{eqnarray}

These equations are completely equivalent to the Maxwell's system
used for the case of particles carrying electric and magnetic charges
(dyons) and thus represent the Generalized Dirac Maxwell (GDM) equations
for inhomogeneous media in quaternionic formulation. The foregoing
analysis is simple, compact and unique one in the sense it is manifestly
covariant under homogeneous Lorentz transformations as well as it
remains invariant under the usual quaternion transformations. It reduces
to the quaternion equations for electromagnetic fields in inhomogeneous
medium described by Kravchenko \cite{key-6} in the absence of magnetic
monopole on dyons. Accordingly it can be discussed for the case of
particles carrying pure magnetic monopole in the absence of electric
charge on dyons.

\end{document}